\documentclass[%
 aip,
 jmp,%
 amsmath,amssymb,
 reprint,%
]{revtex4}

\usepackage{graphicx}
\usepackage{dcolumn}
\usepackage{bm}

\begin{document}
\renewcommand{\baselinestretch}{2}
\small\normalsize

\title{Three-dimensional ghost imaging ladar}

\author{ Wenlin Gong$^{\ast}$, Chengqiang Zhao, Jia Jiao, Enrong Li, Mingliang Chen, Hui Wang, Wendong Xu, and Shensheng Han$^{\ast}$}
\affiliation{Shanghai Institute of Optics and Fine Mechanics,
Chinese Academy of Sciences, Shanghai 201800, China}
\email{gongwl@siom.ac.cn, sshan@mail.shcnc.ac.cn}
\date{\today}

\begin{abstract}
Compared with two-dimensional imaging, three-dimensional imaging is much more advantageous to catch the characteristic information of the target for remote sensing. We report a range-resolving ghost imaging ladar system together with the experimental demonstration of three-dimensional remote sensing with a large field of view. The experiments show that, by measuring the correlation function of intensity fluctuations between two light fields, a three-dimensional map at about 1.0 km range with 25 cm resolution in lateral direction and 60 cm resolution in axial direction has been achieved by time-resolved measurements of the reflection signals.
\end{abstract}

\maketitle

Remote sensing is as one of its most important applications since the laser was invented. The laser system has been mainly used to measure distances to remote targets of interest in the last century. Later, a laser-based rader-type system, called LADAR (Laser Detection And Ranging), was invented to achieve three-dimensional (3D) imaging of the targets of interest and scanning
imaging ladar, at present, becomes a typical imaging system of 3D remote sensing \cite{Albota,Cho,Ahola}. However, because scanning imaging ladar obtains the target's 3D image by scanning the target region point-by-point with a pulsed laser, it is difficult to high-resolution imaging when there is relative motion between the target and imaging ladar \cite{Ahola}.

Ghost imaging (GI), as a nonlocal imaging method, has theoretically and experimentally shown that an unknown object with high lateral resolution can be obtained by using a single-pixel bucket detector and measuring the intensity correlation between two light fields \cite{Pittman,Bennink,Cheng,Angelo,Gatti,Cao,Zhang,Gong,Liu,Gong1,Zhang1,Wang,Erkmen,Shapiro}. And GI has been receiving much more increasing interest for great possible use as a remote-sensing system \cite{Zhang2,Meyers,Zhao,Erkmen1,Shi}. However, all the previous work on GI is focused on two-dimensional imaging in lateral direction.

In remote sensing application, the natural targets we faced with usually are 3D images, where the images comprise of the gray distribution and the information of azimuth-range. For GI, the light fields of two different planes located in the axial correlation depth of light field are always spatially correlated \cite{Liu,Gong1,Zhang1}, so all the targets located in this axial correlation range can be clearly imaged even if the detection plane in the reference path is fixed. However, it is a mixed image, which leads to the loss of the information of the target's range in axial direction. If the reflection signals from the target are recorded by a time-resolved bucket detector, the target's images in different distances can be separately obtained by measuring the intensity correlation of light fields in different time slices. Therefore, when the time-resolved measurement is introduced into the detection of the target's reflection signals, high-resolution imaging both in lateral and axial directions can be realized. In this letter, a practical 3D GI ladar system is proposed and high-resolution 3D remote imaging has been experimentally demonstrated at a distance about 1.0 km range.

Fig. 1 presents the setup of proposed 3D GI ladar. The source, which consists of a 532 nm solid-state pulsed laser with 10 ns pulsed width and a rotating ground glass disk, forms speckle field at stop 1 (field stop). The light emitting from the source propagates an aperture stop 2 and then is divided by a beam splitter (BS) into an object and a reference paths. In the object path, the light
propagates through the objective lens $f_0$ and then to the target. The photons reflected by the target are received by a light concentrator (a telescope with the aperture 140 mm and the focal length 477 mm) and then passes through an interference filter with 1 nm half-width into a photomultiplier tube (PMT). In the reference path, the light goes through the reference lens $f_1$ and then to a CCD camera. In this system, Stop 1 is placed on the conjugate plane of both the target and the CCD camera, which can control the field of view (FOV) on the target plane. The stop 2 is used to control the speckle's transverse size on the target plane and the CCD camera plane. In addition, the PMT connected with a high-speed digitizer of 1 G/s is used to record the light signals reflected from the target in different times, which is equivalent to a time-resolved bucket detector. The interference filter is used for inhibiting background light.

According to GI theory \cite{Bennink,Cheng,Angelo,Gatti}, the target's 3D images can be reconstructed by measuring the correlation function of intensity fluctuations between $B_s$ recorded by the time-resolved bucket detector and the intensity distribution $I(x_r )$ recorded by the CCD camera \cite{Gong2}:
\begin{eqnarray}
\Delta G_s ^{(2,2)} (x_r ) = \left\langle {B_s I(x_r )} \right\rangle  - \left\langle {B_s } \right\rangle \left\langle {I(x_r )} \right\rangle ,s = 1 \cdots m.
\end{eqnarray}
where $\left\langle {\cdots} \right\rangle$ denotes the ensemble average, $\Delta G_s ^{(2,2)} (x_r )$ is the correlation function of intensity fluctuations at the $s$th time slice and $m$ is the number of time slices divided by the high-speed digitizer.

The concrete parameters of 3D GI ladar in the experiments are as follows: the focal length of the lens $f_0$ is 360 mm and the magnification of imaging system in the reference path is 1.75$\times$. The FOV of both the receiving system and the emitting system at ${l_0}'$=1000 m range is about 22 m. According to the parameters of emitting system, the pixel size of the CCD camera in the reference
path is set as 112 $\mu$m$\times$112 $\mu$m, thus the spatial resolution of the pixel size corresponding to the target plane is 178 mm$\times$178 mm at ${l_0}'$=1000 m range.

The performance of 3D GI ladar was first demonstrated by imaging a tower located about ${l_0}'$=570 m away. The target's image, taken by a telescope, is shown in Fig. 2(a). Fig. 2(b) presents a sequence of time-resolved measured signals reflected from the target at one measurement, using the time-resolved bucket detector in the object path. The time resolution of the measured signals is set as 4 ns, which reveals the axial resolution of imaging target is 60 cm. By measuring the correlation function of intensity fluctuations between $B_s$ and $I(x_r )$ in different time slices, the target's tomographic images for the 11th, 12th, 25th and 26th time slices are illustrated in Fig. 2(c$_1$-c$_4$) and Fig. 2(d) shows the target's 3D image when all tomographic GI in different time slices are jointed.

Another demonstration of 3D GI ladar's capability was conducted by imaging a building located about ${l_0}'$=1200 m away. Fig. 3(a) gives the building imaged by a telescope in one evening. By computing the correlation function of intensity fluctuations between $I(x_r )$ and the intensities $B_s$ positioned in the neighborhood of four peaking signals, the corresponding tomographic images of the building are displayed in Fig. 3(c$_1$-c$_4$), respectively. And the building's 3D images are shown in Fig. 3(d) based on GI ladar measurements.

The experimental demonstration of 3D remote sensing with a large field of view was also performed, using the proposed 3D GI ladar system. Fig. 4(a) is the imaging scene located about ${l_0}'$=900 m away and the imaging FOV is approximately 20 m. By choosing the measured signals in some specific time slices correlated with $I(x_r )$, as shown in Fig. 4(b) and Fig. 4(c$_1$-c$_4$), the corresponding partial images of the imaging scene such as trees and houses, can be separately reconstructed. Similar to Fig. 2(d) and Fig. 3(d), Fig. 4 presents 3D images of the imaging scene achieved by the 3D GI ladar system.

In conclusion, a 3D GI ladar system is proposed and experimentally demonstrate that the 3D images of large imaging scenes have been obtained by combining GI method with time-resolved measurements. This technique has a great application prospect in remote sensing.

The work was supported by the Hi-Tech Research and Development Program of China under Grant Project No. 2011AA120101 and No. 2011AA120102.

\newpage
\begin{figure}
\centerline{
\includegraphics[width=14cm]{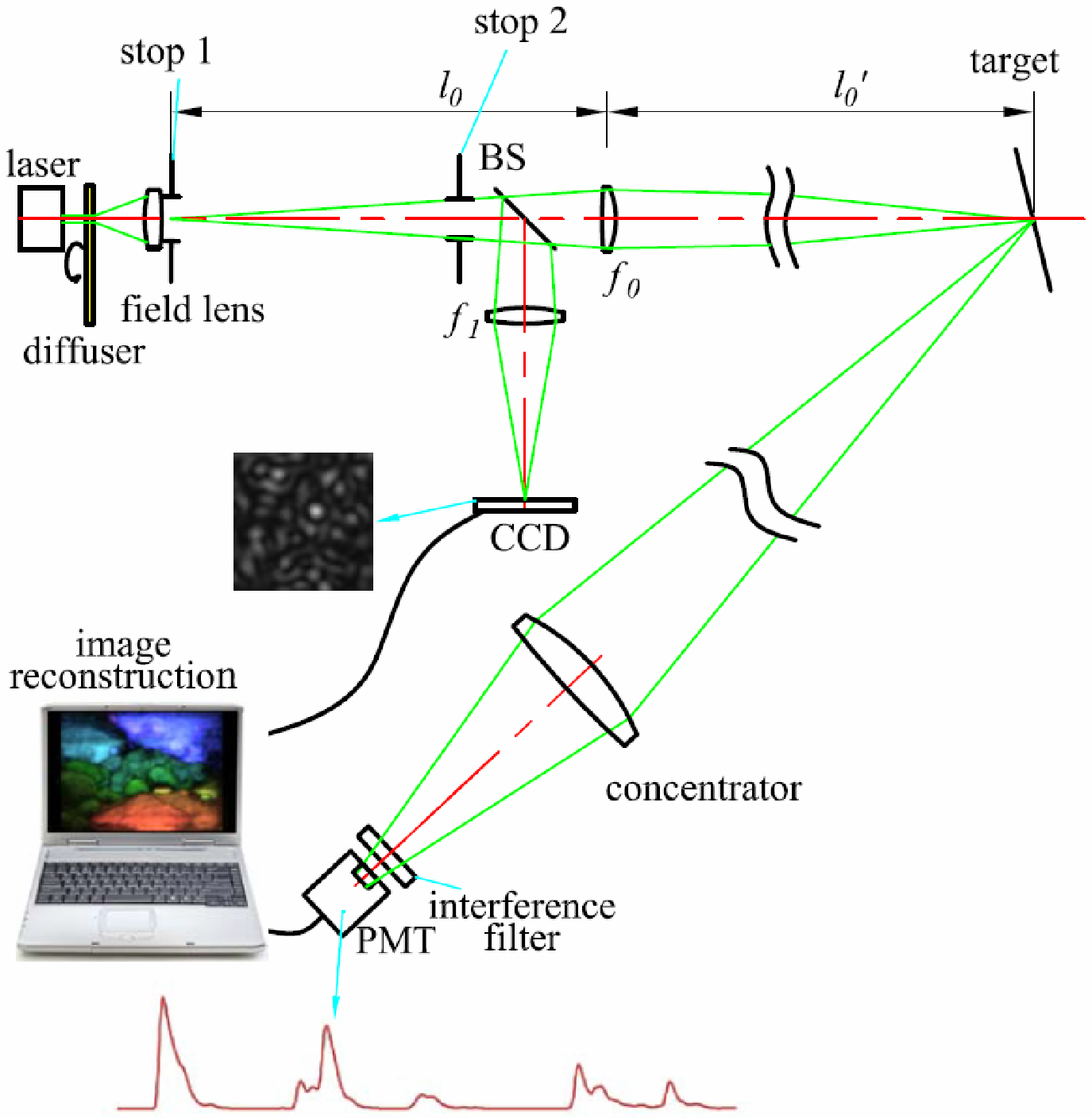}}
\caption{Experimental setup of 3D GI ladar system with pseudo-thermal light.}
\end{figure}

\newpage
\begin{figure}
\centerline{
\includegraphics[width=14cm]{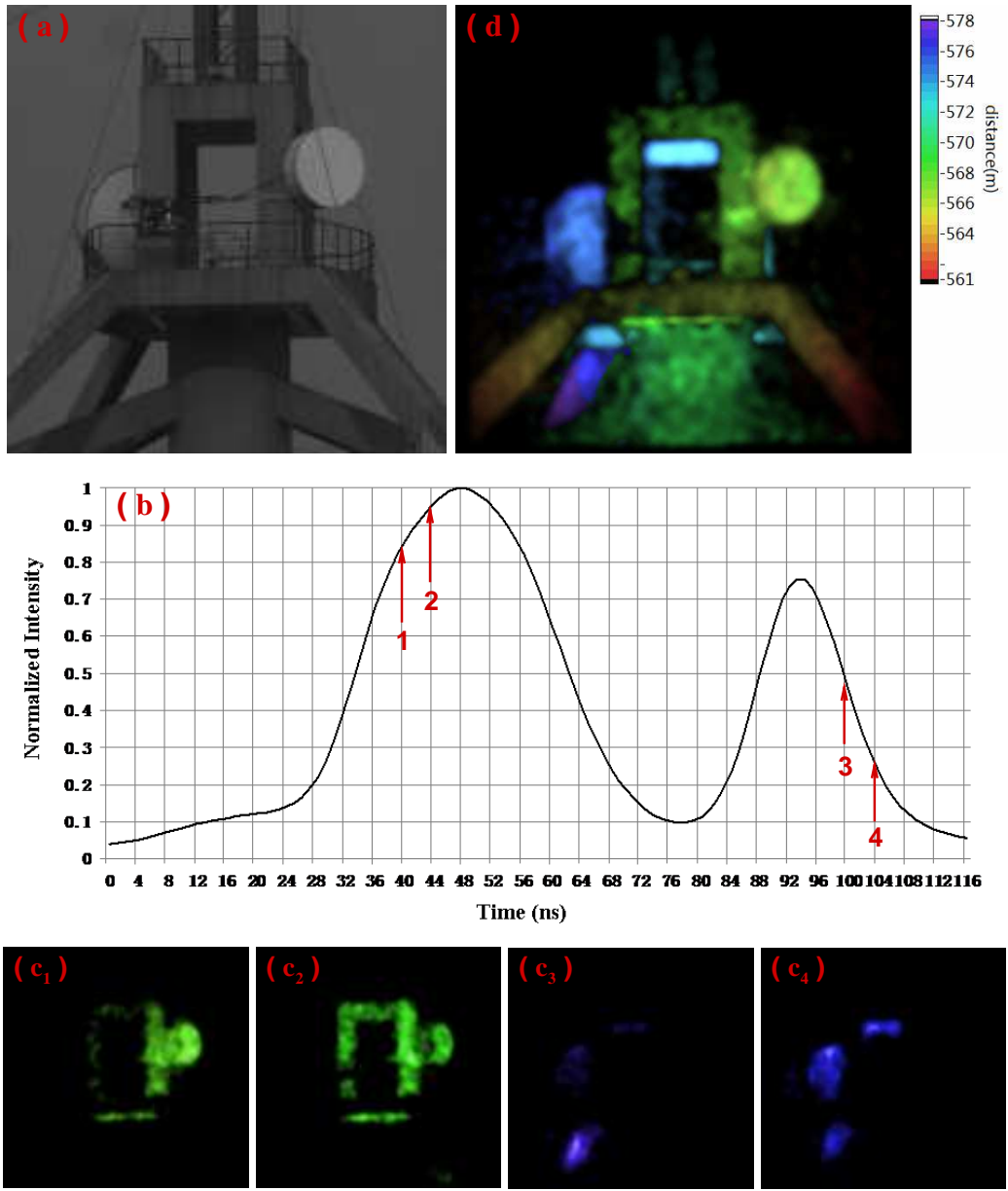}}
\caption{Experimental demonstration results of a tower at about 570 m range, using 10 000 measurements. (a) The original
target imaged by a telescope with the receiving aperture 140 mm; (b) the time-resolved signals reflected from the target at one measurement; (c$_1$-c$_4$) the tomographic images of the target reconstructed by GI method, corresponding to the labeled time slices of Fig. 2(b); (d) the target's 3D images obtained by GI ladar.}
\end{figure}

\newpage
\begin{figure}
\centerline{
\includegraphics[width=14cm]{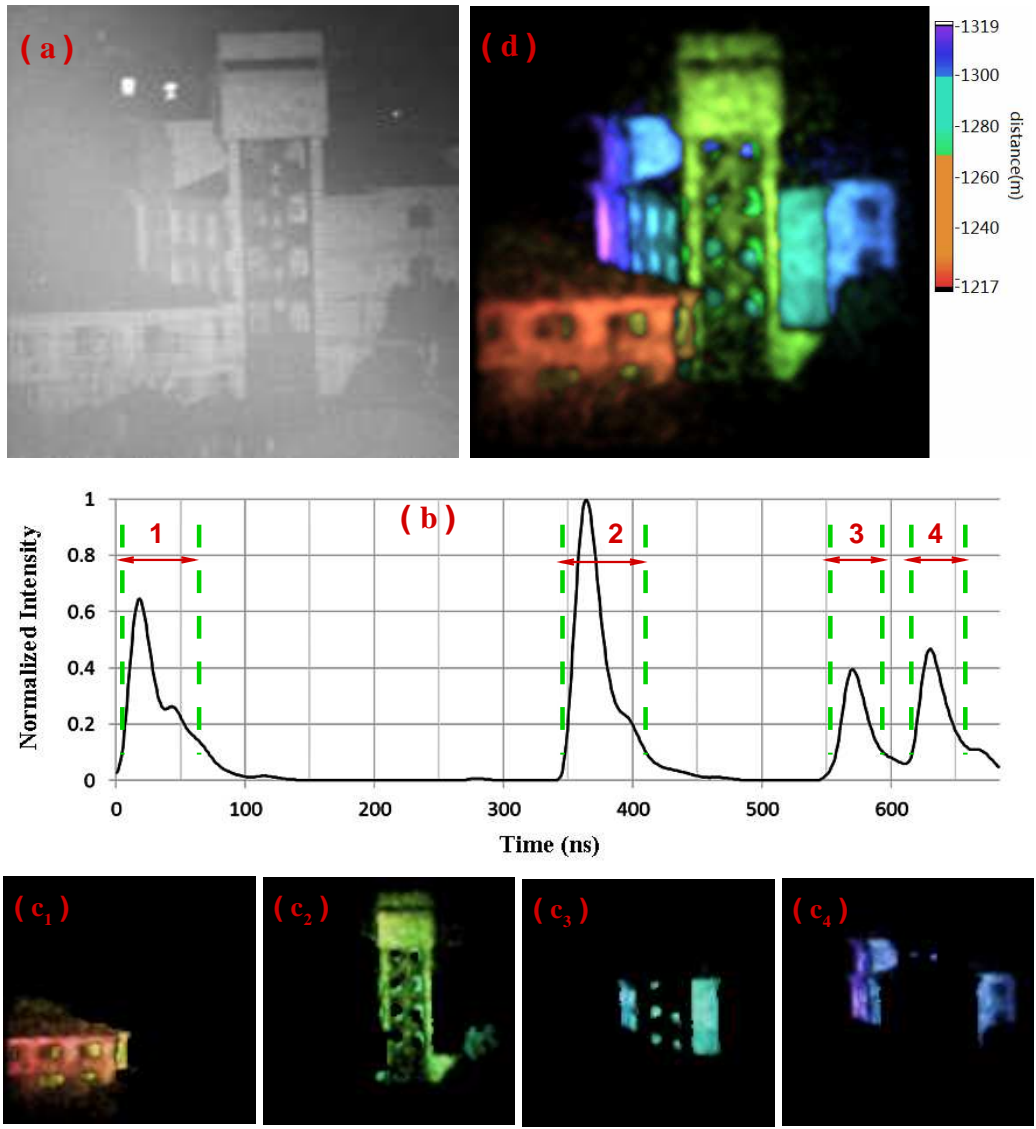}}
\caption{Experimental reconstruction results of imaging a building located about ${l_0}'$=1200 m away (averaged 10 000 measurements), the captions are the same as Fig. 2.}
\end{figure}

\newpage
\begin{figure}
\centerline{
\includegraphics[width=14cm]{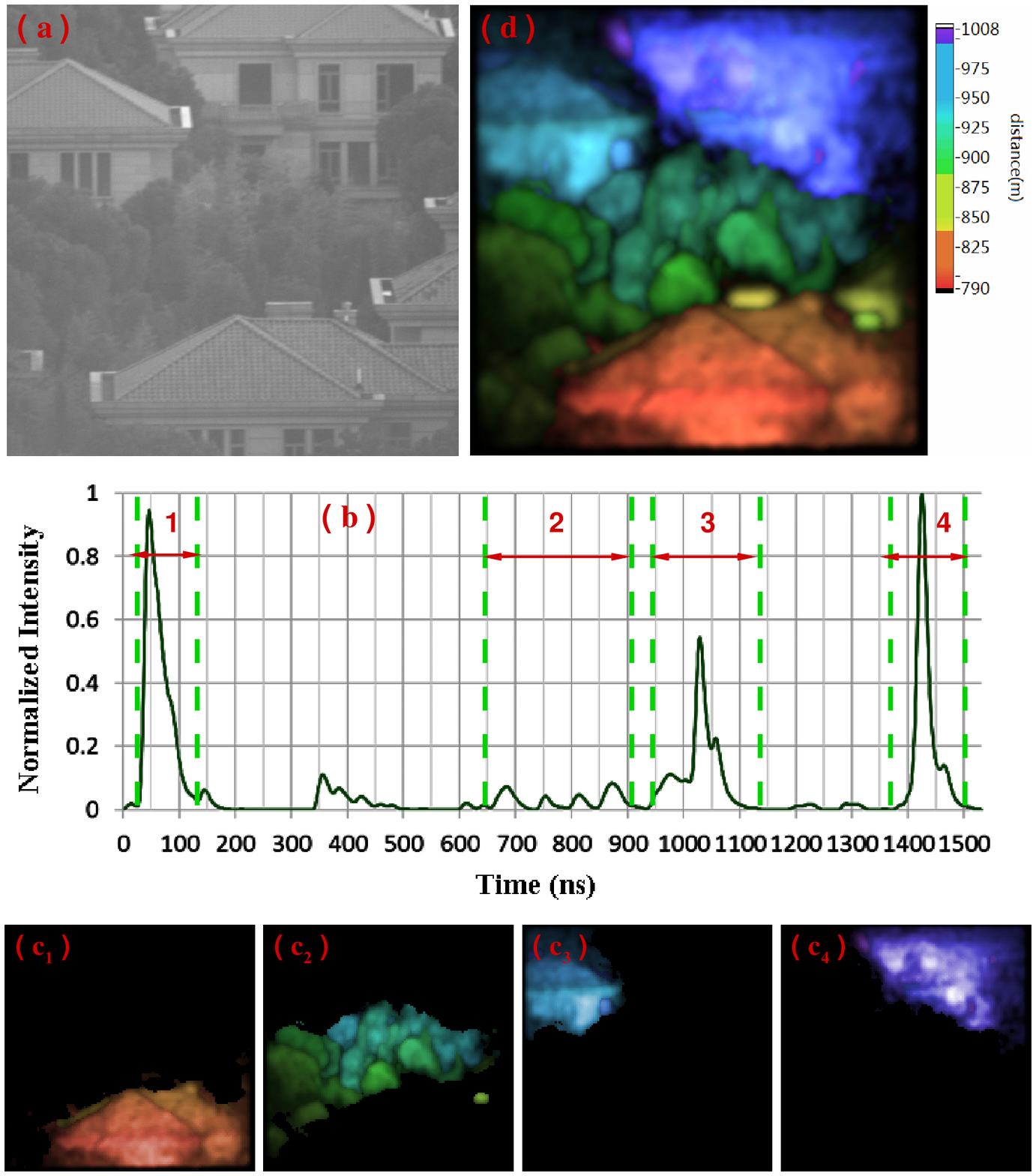}}
\caption{Experimental demonstration results of a large imaging scene located about ${l_0}'$=900 m away, the captions are the same as Fig. 2 and Fig. 3.}
\end{figure}


\begin{thebibliography}{99}
\bibitem{Albota} M. A. Albota, B. F. Aull, D. G. Fouche, R. M. Geinriches, D. G. Kocher, R. M. Marino, J. G. Moony, N. R. Newbury, M. E. O'Brien, B. E. Player, B. C. Willard, and J. J. Zayhowski, Lincoln Lab. J. \textbf{13}, 351-370 (2002).
\bibitem{Cho} P. Cho, H. Anderson, R. Hatch, and P. Ramaswami, Lincoln Lab. J. \textbf{16}, 147-164 (2006).
\bibitem{Ahola} R. Ahola, T Heikkinen, and M. Manninen, Proc. Int. Conf. Image Processing and Pattern Recngnition, (1985), p. 139.
\bibitem{Pittman} T. B. Pittman, Y. H. Shih, D. V. Strekalov, and A. V. Sergienko, Phys. Rev. A. \textbf{52}, R3429 (1995).
\bibitem{Bennink} R. S. Bennink, S. J. Bentley, and R. W. Boyd, Phys. Rev. Lett. \textbf{89}, 113601 (2002).
\bibitem{Cheng} J. Cheng and S. Han, Phys. Rev. Lett. \textbf{92}, 093903 (2004).
\bibitem{Angelo} M. D. Angelo and Y. H. Shih, Laser Phys. Lett. \textbf{2}, 567-596 (2005).
\bibitem{Gatti} A. Gatti, M. Bache, D. Magatti, E. Brambilla, F. Ferri, and L. A. Lugiato, J. Mod. Opt. \textbf{53}, 739-760 (2006).
\bibitem{Cao} D. Z. Cao, J. Xiong, K. Wang, Phys. Rev. A \textbf{71}, 013801 (2005).
\bibitem{Zhang} D. Zhang, Y-H. Zhai, L-A. Wu and X-H. Chen, Opt. Lett. \textbf{30}, 2354-2356 (2005).
\bibitem{Gong} W. Gong, P. Zhang, X. Shen, and S. Han, Appl. Phys. Lett. \textbf{95}, 071110 (2009).
\bibitem{Liu} H. Liu, and S. Han, Opt. Lett. \textbf{33}, 824-826 (2008).
\bibitem{Gong1} W. Gong, S. Han, J. Opt. Soc. Am. B \textbf{27}, 675-678 (2010).
\bibitem{Zhang1} P. Zhang, W. Gong, X. Shen, and S. Han, Opt. Lett. \textbf{34}, 1222-1224 (2009).
\bibitem{Wang} C. Wang, D. Zhang, Y. Bai, and B. Chen, Phys. Rev. A. \textbf{82}, 063814 (2010).
\bibitem{Erkmen} B. I. Erkmen, and J. H. Shapiro, Adv. Opt. Photon. \textbf{2}, 405-450 (2010).
\bibitem{Shapiro} J. H. Shapiro, and R. W. Boyd, Quantum Inf. Process. \textbf{11}, 949-993 (2012).
\bibitem{Zhang2} P. Zhang, W. Gong, X. Shen, and S. Han, Phys. Rev. A \textbf{82}, 033817 (2010).
\bibitem{Meyers} R. E. Meyers, K. S. Deacon, and Y. Shih, Appl. Phys. Lett. \textbf{98}, 111115 (2011).
\bibitem{Zhao} C. Zhao, W. Gong, M. Chen, E. Li, H. Wang, W. Xu, and S. Han, Appl. Phys. Lett. \textbf{101}, 141123 (2012).
\bibitem{Erkmen1} B. I. Erkmen, J. Opt. Soc. Am. A \textbf{29}, 782-789 (2012).
\bibitem{Shi} D. Shi, C. Fan, P. Zhang, J. Zhang, H. Shen, C. Qiao, and Y. Wang, Opt. Express \textbf{20}, 27992-27998 (2012).
\bibitem{Gong2} W. Gong, and S. Han, Phys. Lett. A. \textbf{374}, 1005-1008 (2010).
\end{thebibliography}
\end{document}